\documentclass[aps,prb,twocolumn,groupedaddress,showpacs]{revtex4}

\usepackage{amsmath}
\usepackage{amssymb}
\usepackage{epsfig}
\newcommand {\dr}{{\mathrm d}\mathbf{r}}

\newcommand {\dd}{{\mathrm d}}
\newcommand {\rr}{\mathbf{r}}

\newcommand {\im}{\mathrm i}

\begin{document}


\title{Pair correlation functions and phase separation in a two component
point Yukawa fluid}


\author{P. Hopkins}
\email[]{Paul.Hopkins@bristol.ac.uk}
\author{A.J. Archer}
\email[]{Andrew.Archer@bristol.ac.uk}
\author{R. Evans}
\affiliation{H.H. Wills Physics Laboratory,
University of Bristol, Bristol BS8 1TL, UK}


\date{\today}

\begin{abstract}
We investigate the structure of a binary mixture of particles interacting
via purely repulsive (point) Yukawa pair potentials with a common inverse
screening length $\lambda$. Using the
hyper-netted chain closure to the Ornstein-Zernike equations, we find that for a
system with `ideal' (Berthelot mixing rule) pair potential parameters for the
interaction between unlike species, the asymptotic decay of the total
correlation functions crosses over from monotonic to damped oscillatory on
increasing the fluid total density at fixed composition. This gives rise to a
Kirkwood line in the phase diagram. We also consider a `non-ideal' system, in
which the Berthelot mixing rule is multiplied by a factor $(1+\delta)$. For any
$\delta>0$ the system exhibits fluid-fluid phase separation and remarkably the
ultimate decay of the correlation functions is now monotonic for {\em all}
(mixture) state points. Only in the limit of vanishing concentration of either
species does one find oscillatory decay extending to $r = \infty$. In the
non--ideal case the simple random phase approximation provides a good
description of the phase separation and the accompanying Lifshitz line.
\end{abstract}


\maketitle

\section{Introduction}
\label{sec:intro}

A large class of fluids can be described generically as ``big charged
particles immersed in a neutralising medium of lighter particles''. Examples
include charged colloidal suspensions\cite{Hansen} and
dusty plasmas.\cite{piel:melzer2002} A simple
model for such systems describes the effective interaction between the
bigger particles in terms of a Yukawa (screened Coulomb) pair potential $\phi(r)
\propto \exp(-\lambda r)/r$. The effects of the screening due to the
neutralising medium are incorporated via the screening parameter $\lambda$. Such
a Yukawa potential arises in, for example, the linearised Poisson--Boltzmann or
Derjaguin--Landau--Verwey--Overbeek theories for the effective potential between
spherical
charged colloids in solution.\cite{Hansen} The effect of the neutralising
medium on the effective potential involves more than the screening effect, as
described by the parameter $\lambda$. There is an additional effect of charge
renormalization, whereby the amplitude of the effective potential $\phi(r)$ is
not, as one might perhaps expect from a linear treatment, proportional to $Z^2$,
where $Z$ is the charge on the big particles (colloids), rather the amplitude of
$\phi(r)$ is proportional to $\bar{Z}^2$, where $\bar{Z}<Z$ is the renormalised
charge.\cite{Hansen}

In the present paper we are concerned with a simple model of a {\em binary}
mixture of big charged particles, with both species carrying charges of the same
sign, immersed in a neutralising background medium.
For a general binary mixture of point Yukawa particles the pair potentials are
dependent on six species specific parameters. These are the dimensionless
coupling parameters, $M_{ij}$, and the screening parameters, $\lambda_{ij}$, for
$i,j=1,2$. We assume that both species experience the same screening,
$\lambda_{ij}=\lambda$, determined by the background medium (solvent), but that
they have different coupling strengths. We can consider the coupling parameters
to be proportional to the product of the effective charges of the species. Thus
we define the pair potential as
\begin{equation}
\phi_{ij}(r)=\frac{M_{ij}\epsilon\exp(-\lambda r)} {\lambda r},
\label{eq:pp1}
\end{equation}
where $\epsilon$ denotes the overall energy scale, and all three potentials are
repulsive: $M_{ij}>0$. As usual it is assumed that the inter-species parameters
are related to those for like particles.\cite{HanMcD} The Berthelot mixing rule
sets $M_{12}=\sqrt{M_{11}M_{22}}$. This choice could correspond to ions with
charge $\bar{Z}_i$, with $\bar{Z}_1, \bar{Z}_2>0$, immersed in a medium with an
inverse screening length $\lambda$. In order to generalise this mixing rule we
introduce a non-ideality parameter $\delta>0$ such that
$M_{12}=(1+\delta)\sqrt{M_{11}M_{22}}$. The case $\delta=0$ clearly corresponds
to the ideal system. The non-ideal case $\delta\neq0$ can be viewed as arising
from charge screening effects in the double layer of condensed counter ions on
the surface of the particles. The double layer leads to an effective
renormalisation of the particle charge and strongly affects the particle
interactions.\cite{Hansen} One should expect that for some positive $\delta$
that the energy penalty incurred for unlike species to be neighbours should lead
to fluid-fluid demixing at high densities; such behaviour is found in models of
soft-core fluids where positive non-additivity gives rise to
demixing.\cite{Archer02, Archer01, Finken, Louis}

Binary mixtures of H$^+$ and He$^{++}$ in a neutralising medium are expected to
phase separate at temperatures and pressures of astrophysical interest
-- see Ref.~\onlinecite{Baus}. In this system the phase separation is
thought to be due to charge neutralisation being less efficient in the mixture
than in the pure phases. We believe that some of the complex screening effects
associated with such systems may be incorporated in a simple model such as ours,
via the parameter $\delta$. Non--ideal charge renormalisation effects may also
be present in binary suspensions of colloids. Charge renormalisation may be
affected by the local concentrations of the different species of colloids and
counterions in very subtle ways leading to the possibility that the amplitude of
$\phi_{12}(r)$ may not be proportional to $\bar{Z}_1 \bar{Z}_2$. Such effects
would be mimicked using our simple
model. It may be the case that in some colloidal fluid mixtures $\delta>0$, or
it may be that in other cases $\delta<0$, i.e.~non-ideal charge renormalization
effects may favour mixing. However, in the present paper we investigate only the
cases $\delta=0$ and $\delta >0$. Using the accurate hyper-netted chain (HNC)
approximation we find that for the case $\delta=0$ the total pair correlation
functions $h_{ij}(r)$ exhibit crossover from monotonic to exponentially damped
oscillatory asymptotic decay, $r\rightarrow\infty$, on increasing the total
density of
the fluid mixture at fixed composition. This scenario is equivalent to that
observed in the one-component point Yukawa fluid (OCY).\cite{Hopkins} However,
in the non-ideal case, an {\em infinitesimal} positive $\delta$ can give rise to
fluid-fluid phase separation at a sufficiently large total density. Moreover we
find that the fluid structure is changed profoundly from that pertaining to
$\delta=0$, i.e.~for all thermodynamic state points, apart from the limits of
pure species 1 and 2, the ultimate, $r\rightarrow\infty$, decay of correlations
is monotonic. We find that for the particular choice $\delta=0.1$ the very
simple random phase approximation (RPA) provides a good account of the
fluid-fluid binodal and spinodal obtained from the HNC but a poor account of the
detailed behaviour of the correlation functions.

The paper is arranged as follows:
In Secs.~\ref{sec:OZ_eq} and \ref{sec:asymp_decay} we remind readers of the
HNC and RPA integral equations and
some basic results from the theory of the asymptotic decay of pair correlation
functions in binary mixtures. Sec.~\ref{sec:results_g} describes results for
$h_{ij}(r)$ and for
the poles of the Fourier transforms $\hat{h}_{ij}(q)$ obtained from numerical
solutions of the HNC closure approximation. Within the RPA we are able to
calculate the corresponding poles analytically and we compare these results with
those from the HNC. The pole analysis enables us to determine the behaviour of
$h_{ij}(r)$ at intermediate range, as well as at longest range,
$r\rightarrow\infty$. Particular attention is paid to the behaviour of the
correlation functions in the limit where the density of species 2, $\rho_2
\rightarrow 0$. In Sec.~\ref{sec:phase_diag} we present phase diagrams,
in the composition,
total density plane, along with Lifshitz lines for the partial structure
factors, obtained from both the HNC and RPA.
We draw some conclusions in Sec.~\ref{sec:discussion}.

\section{Closure of the OZ equations}
\label{sec:OZ_eq}

Our starting point for determining the fluid structure is the mixture
Ornstein-Zernike (OZ) equation\cite{HanMcD} which relates the total correlation
functions, $h_{ij}(r)=g_{ij}(r)-1$, where $g_{ij}(r)$ are the radial
distribution functions, to a set of pair direct correlation functions,
$c_{ij}(r)$:
\begin{equation}
h_{ij}(r_{12})=c_{ij}(r_{12})
+\sum_{k=1}^2\rho_{k}\int \dd^3r_3 c_{ik}(r_{13})h_{kj}(r_{32}),
\label{eq:oz1}
\end{equation}
where $r_{ij}=|\rr_i-\rr_j|$ and $\rho_k$ is the bulk density of species $k$.
These equations can be viewed as defining the pair direct correlation functions.
In order to determine the fluid structure a second relation, or closure, is
required.

The simplest closure germane to the present model is the random phase
approximation (RPA): $c_{ij}^{RPA}=-\beta\phi_{ij}(r)$ with $\beta=(k_BT)^{-1}$,
which is strictly valid only for $r\rightarrow\infty$. Although this
approximation is inadequate for hard-core model systems, it has been shown that
the RPA becomes accurate for some soft-core systems at intermediate densities
and exact at high densities.\cite{Archer02, Archer01, Finken, Louis, Likos1,
Likos2, Lang} An important advantage of the RPA is that
it does provide an analytical solution for correlation functions and for
thermodynamic properties which may provide valuable physical insight into fluid
behaviour.

A more accurate approximation is the hyper-netted chain (HNC) approximation.
The exact closure of the OZ equations can be expressed as:
\begin{equation}
g_{ij}(r)=\exp(-\beta\phi_{ij}(r)+h_{ij}(r)-c_{ij}(r)-b_{ij}(r)), 
\label{eq:hnc}
\end{equation}
where $-b_{ij}(r)$ is an unknown bridge function.\cite{HanMcD} The HNC simply
sets this bridge function to zero for all $r$. It is found to be accurate for
long-ranged or soft-core potentials,\cite{HanMcD, Baus, Likos1} although it may
fail in the neighbourhood of a spinodal. For the one-component (point) Yukawa
fluid it is known that the HNC is remarkably accurate for
small coupling parameters.\cite{Hopkins, Daughton} In order to determine the
correlation functions within the HNC we use a standard iterative procedure. In
what follows, we fix the reduced temperature $T^*=(\beta \epsilon)^{-1}$ so that
the state of the system is determined by the total density,
$\rho=\rho_1+\rho_2$, and the species concentrations $x_i=\rho_i/\rho$.

\section{Asymptotic decay of correlation functions}
\label{sec:asymp_decay}

There are two procedures for determining the asymptotic, $r\rightarrow\infty$,
behaviour of the total correlation functions. One is to examine directly the
numerical solutions for $h_{ij}(r)$. The alternative method is to input the
direct correlation functions (in the present case from either the RPA or HNC
closures) into the set of of OZ equations (\ref{eq:oz1}) and perform an
asymptotic analysis. The OZ equations can be solved formally in Fourier space
and the solution written as
\begin{equation}
\hat{h}_{ij}(q)=\frac{N_{ij}(q)}{D(q)},
\label{eq:oz2}
\end{equation}
where $\hat{h}_{ij}(q)$ denotes the three-dimensional Fourier transform of
$h_{ij}(r)$. The three functions share the same denominator
\begin{equation}
D(q)=[1-\rho_{1}\hat{c}_{11}(q)][1-\rho_{2}\hat{c}_{22}(q)]
-\rho_{1}\rho_{2}\hat{c}_{12}(q)^{2},
\label{eq:den}
\end{equation}
but the numerators are dependent on the indices:
\begin{eqnarray}
N_{11}(q)&=&\hat{c}_{11}(q)+\rho_{2}
[\hat{c}_{12}(q)^{2}-\hat{c}_{11}(q)\hat{c}_{22}(q)],\nonumber\\
N_{22}(q)&=&\hat{c}_{22}(q)+\rho_{1}
[\hat{c}_{12}(q)^{2}-\hat{c}_{11}(q)\hat{c}_{22}(q)],\label{eq:oznum}\\
N_{12}(q)&=&N_{21}(q)=\hat{c}_{12}(q).\nonumber
\end{eqnarray}
From the inverse Fourier transform it follows that:
\begin{equation}
rh_{ij}(r)=\frac{1}{2\pi^{2}} \int_{0}^{\infty} \dd q q \sin(qr)\hat{h}_{ij}(q).
\label{eq:hrint}
\end{equation}
Using Eq.(\ref{eq:oz2}) and assuming that the singularities of
$\hat{h}_{ij}(q)$ for
the present Yukawa systems are simple poles we are able to proceed via the
residue theorem.\cite{Evans} Performing contour integration around a semicircle
in the upper half of the complex $q$ plane, we write the total correlation
functions as a sum of contributions from the poles enclosed:
\begin{equation}
rh_{ij}(r)=\sum_{n}A_{n}^{ij}\exp(\im q_{n}r),
\label{eq:hrsum}
\end{equation}
where $q_n$ satisfies $D(q_n)=0$ and $A_{n}^{ij}$ is the amplitude associated
with the pole $q_{n}$. This amplitude is related to the residue $R_n^{ij}$ of
$qN_{ij}(q)/D(q)$ by $A_n^{ij}=R_n^{ij}/2\pi$. The poles are either purely
imaginary, $q=\im\alpha_0$, or occur as a conjugate complex pair
$q=\pm\alpha_1+\im\tilde{\alpha}_0$. \cite{Evans}

In general there are an infinite number of poles and contributions from many of
these are required to account for the behaviour of $h_{ij}(r)$ at small $r$.
However, the ultimate, $r\rightarrow\infty$, decay of $h_{ij}(r)$ is determined
by the pole that gives the slowest exponential decay, i.e.~the pole with the
smallest imaginary part. This is referred to as the leading order pole. If the
leading order pole is purely imaginary then $rh_{ij}(r)$ decays exponentially,
$rh_{ij}(r)\sim A_{ij}\exp(-\alpha_{0}r)$, as $r\rightarrow \infty$. On the
other hand, if the leading order poles are a conjugate complex pair, then the
sum of contributions from this pair of complex poles gives damped oscillatory
decay, $rh_{ij}(r)\sim
2\tilde{A}_{ij}\exp(-\tilde{\alpha}_{0}r)\cos(\alpha_{1}r-\tilde{\theta}_{ij})$,
where $\tilde{A}_{ij}$ and $\tilde{\theta}_{ij}$ denote the amplitude and phase
respectively.\cite{Evans}

Note that whereas the wavelength $2\pi/\alpha_1$ and the decay lengths
$\alpha_0^{-1}$ or $\tilde{\alpha}_0^{-1}$ are the same for all $h_{ij}(r)$, the
amplitudes and phases do depend on the indices $ij$.\cite{Evans} However,
general considerations demand $A_{12}^2=A_{11}A_{22}$ or
$\tilde{A}_{12}^2=\tilde{A}_{11}\tilde{A}_{22}$ and
$2\tilde{\theta}_{12}=\tilde{\theta}_{11}+\tilde{\theta}_{22}$.\cite{Evans} In
the next section we shall employ the RPA and the HNC closures to investigate the
ultimate, $r\rightarrow\infty$, decay of $h_{ij}(r)$ and the behaviour of
$h_{ij}(r)$ at intermediate distances $r$. Note that the values of the
amplitudes are relevant in determining which pole or conjugate complex pair
provides the dominant contribution to $h_{ij}(r)$ in the intermediate regime.

\section{Results for Pair Correlation Functions}
\label{sec:results_g}

\begin{figure}
\includegraphics[width=8cm]{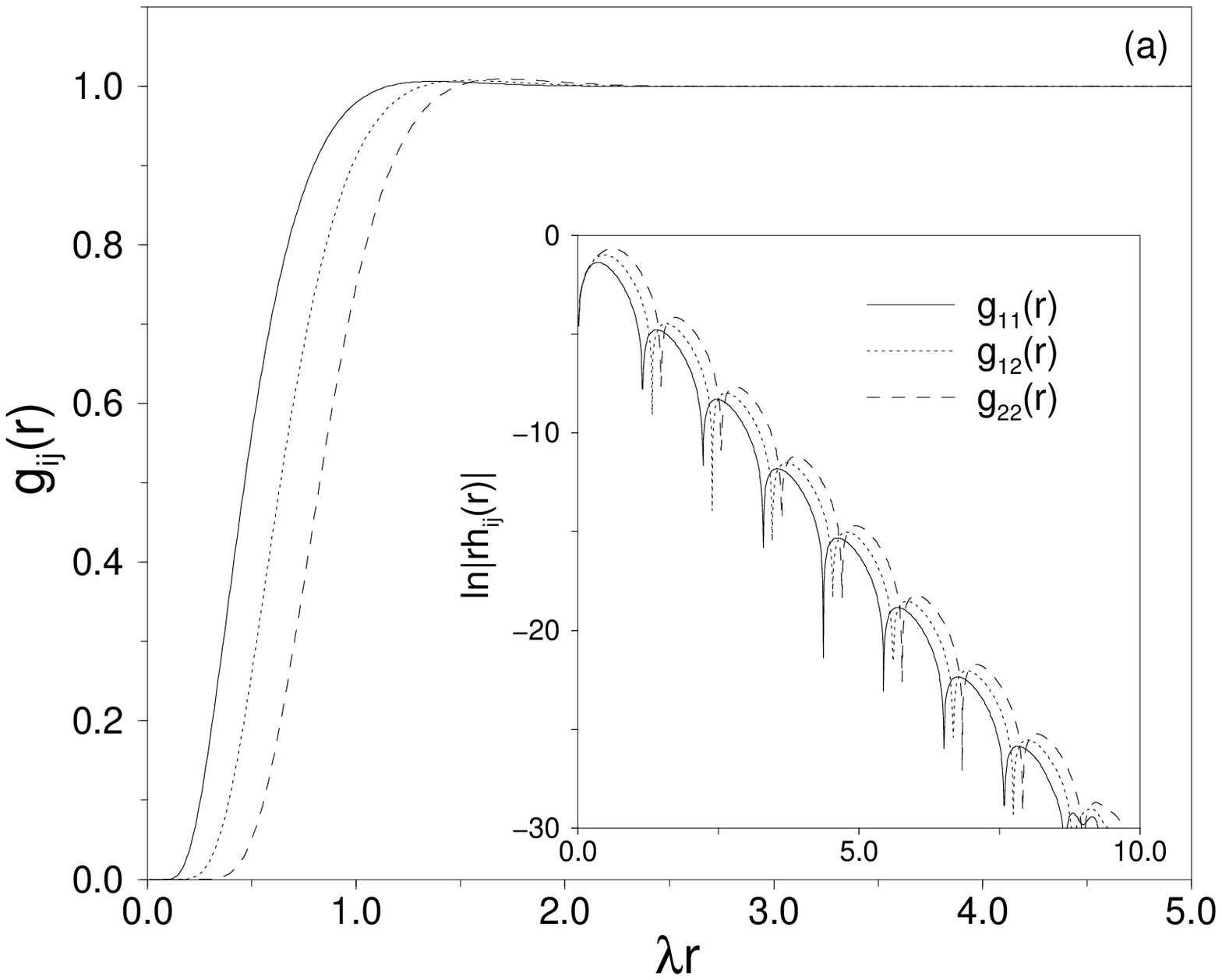}
\includegraphics[width=8cm]{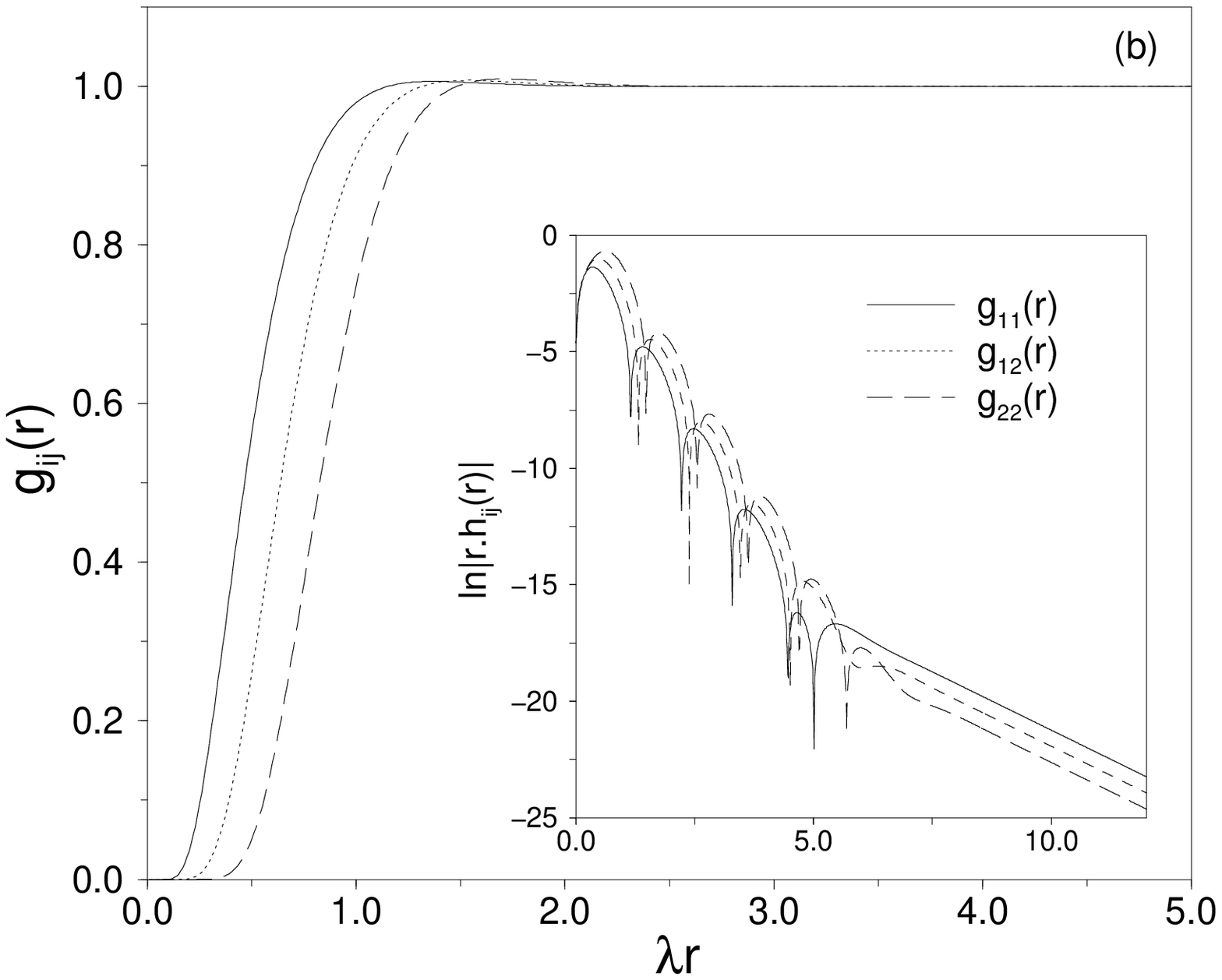}
\includegraphics[width=8cm]{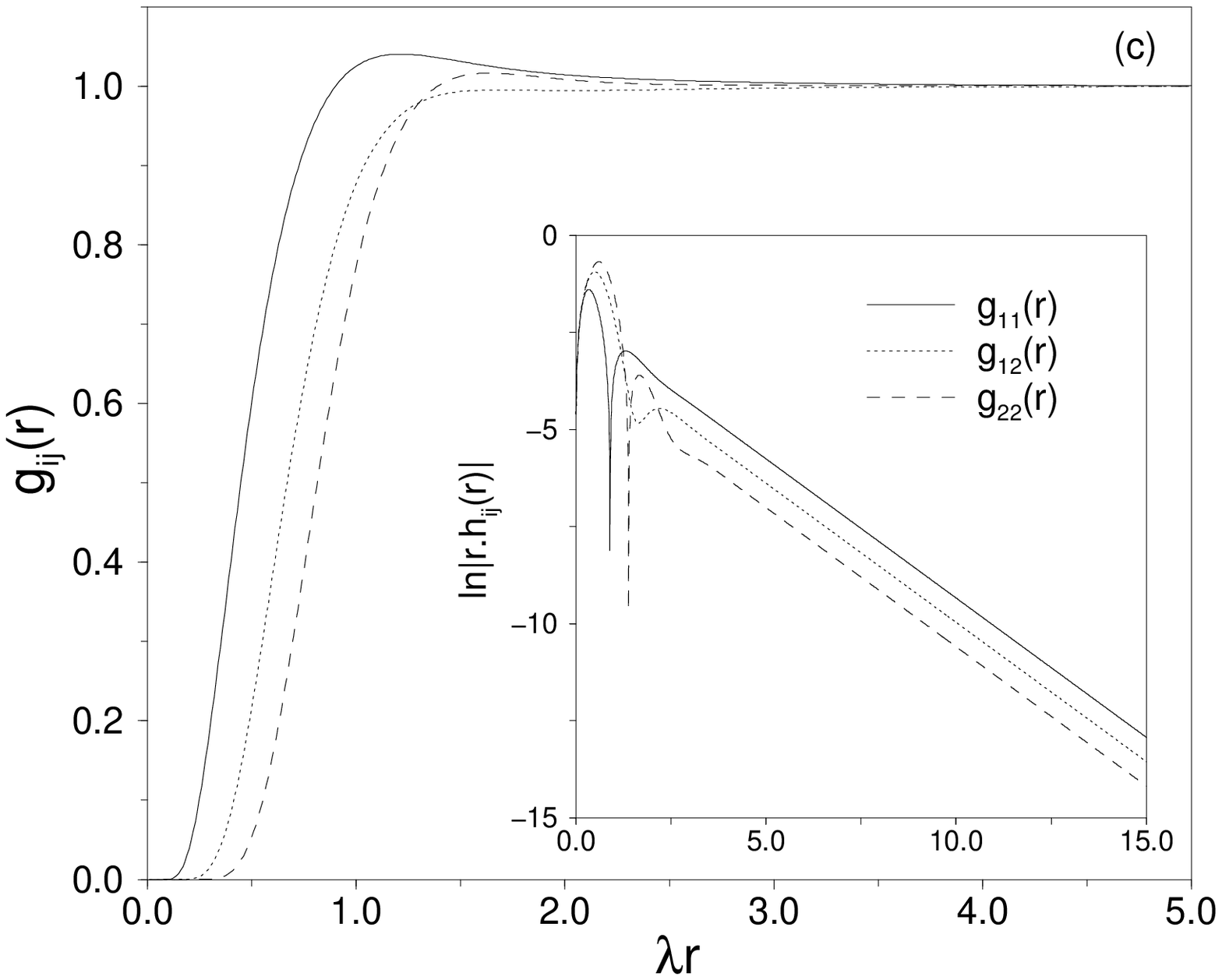}
\caption{\label{fig:rdfs}
Pair correlation functions for the state point $\rho\lambda^{-3}=0.5$, $x_2=0.5$
and $T^*=1$, calculated from the HNC closure, using the coupling parameter ratio
$M_{22}/M_{11}=4$. Main figures show radial distribution functions, $g_{ij}(r)$,
and insets show $\ln|rh_{ij}(r)|$ versus $\lambda r$. a) $\delta=0$.
Exponentially damped oscillations extend to infinity. b) $\delta=10^{-5}$. For
$\lambda r\gtrsim8$, $rh_{ij}(r)$ exhibit monotonic (exponential) decay. c)
$\delta=0.1$. Monotonic decay now develops for $\lambda r \gtrsim 3$.}
\end{figure}

In Fig.~\ref{fig:rdfs} we display the radial distribution functions $g_{ij}(r)$,
obtained from the HNC closure, for the state point $\rho\lambda^{-3}=0.5$,
$x_2=0.5$ and $T^*=1$. The coupling parameters are fixed at $M_{11}=1$, and
$M_{22}=4$. Fig.~\ref{fig:rdfs}(a) corresponds to ideal (Berthelot) mixing,
$\delta=0$, Fig.~\ref{fig:rdfs}(b) has $\delta=10^{-5}$, weak non-ideality,
while Fig.~\ref{fig:rdfs}(c) has $\delta=0.1$. In all cases the $g_{ij}(r)$
appear structureless but the expanded scales of the insets reveal the nature of
the intermediate and long range decay of the pair correlation functions. For
$\delta=0$, $rh_{ij}(r)$ shows exponentially damped oscillatory decay extending
to arbitrarily large separations $r$. By contrast, for $\delta=10^{-5}$, there
are several oscillations at small $r$ and $rh_{ij}(r)$ decays monotonically
(exponentially) for $\lambda r\gtrsim8$. For $\delta=0.1$ the monotonic decay
extends from $\lambda r\simeq 3$. Clearly the choice of $\delta$ has a profound
influence on the behaviour of the pair correlation functions. If one fixes the
concentration at $x_2=0.5$ and reduces the total density $\rho$ one finds that
for $\delta=0$ there is a crossover from the exponentially damped oscillatory
behaviour of $rh_{ij}(r)$ shown in Fig.~\ref{fig:rdfs}(a) to monotonic
(exponential) decay similar shown to that in Fig.~\ref{fig:rdfs}(c), near
$\rho\lambda^{-3}\simeq0.05$. Such crossover is found in the OCY on reducing the
density at fixed $T^*$.\cite{Hopkins} The scenario is different for $\delta>0$.
On reducing the density at fixed $x_2=0.5$ the $h_{ij}(r)$ are similar to those
for $\rho\lambda^{-3}=0.5$; the decay remains monotonic. Similar features are
found for other values of the concentration provided $x_1,x_2\neq0$. If $x_2=0$
or $1$ then one recovers the OCY which exhibits crossover as mentioned above.

In order to understand these results emanating from the full numerical solution
of the HNC closure we turn to the asymptotic (pole) analysis of
Sec.~\ref{sec:asymp_decay}. It is
convenient to begin with the simple RPA treatment before discussing the more
sophisticated HNC results.

\subsection{Poles in the RPA}
\label{subsec:RPA_poles}

The advantage of the RPA is that the pair direct correlation functions and their
Fourier transforms, $\hat{c}_{ij}(q)$, are given analytically. This means that
the poles can be determined analytically. Using the definition of the RPA it
follows from Eq.~(\ref{eq:pp1}) that
\begin{equation}
\hat{c}_{ij}^{RPA}(q)=-\frac{4\pi M_{ij}}{\lambda T^*}
\frac{1}{(\lambda^2+q^2)}.
\label{eq:rpadcf}
\end{equation}
Substituting this form into Eq.~(\ref{eq:den}) we can solve for the zeros of
$D(q)$, i.e.~the poles, $q_n$, at each state point. We find that within the RPA
there are only two poles. Both are purely imaginary and are given by
\begin{equation}
\alpha_{0}=\sqrt{\frac{2\pi\rho}{\lambda T^{*}}
\left(M_0\pm\sqrt{M_0^2+M_{\delta}}\right)+\lambda^{2}}.
\label{eq:rpasol}
\end{equation}
where we have introduced $M_0=x_1M_{11}+x_2M_{22}$ and
$M_{\delta}=4x_1x_2M_{11}M_{22}(2+\delta)\delta$.

\begin{figure}
\includegraphics[width=8cm]{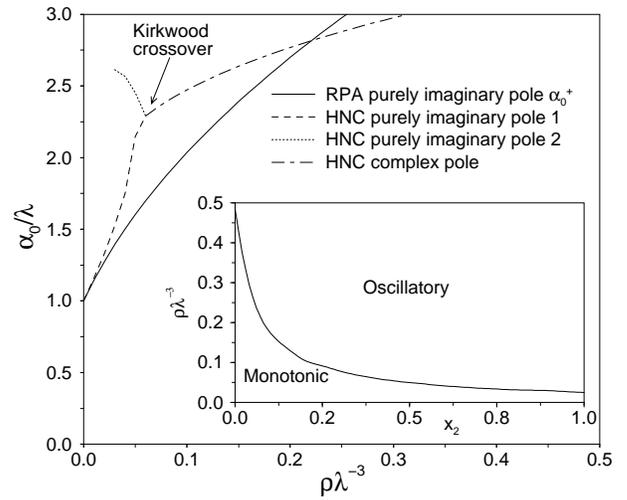}
\caption{\label{fig:poles1}
The imaginary components of leading order poles, $\alpha_0$, calculated along a
path in the phase diagram of increasing density, $\rho$, for fixed $x_2=0.5$ and
$T^*=1$. $\delta=0$, corresponding to an `ideal' mixture. The RPA solution,
Eq.~(\ref{eq:rpasol}), consists of a single, purely imaginary pole,
$\alpha_0^+$, which increases steadily with density from $\alpha_0/\lambda=1$.
Within the HNC closure, at low densities $\rho$, the two leading order poles are
both purely imaginary. As the density is increased the first imaginary pole
ascends and the  second imaginary pole descends and, at the Kirkwood crossover
point,
coalesce becoming a conjugate complex pair $q=\pm\alpha_1+i\alpha_0$. Increasing
the density further  both the real and imaginary parts
of the complex poles increase.
Thus the Kirkwood point marks the boundary between monotonic (at small
$\rho\lambda^{-3}$) and damped oscillatory asymptotic decay. The inset shows the
Kirkwood line, separating the two types of decay, plotted in the
concentration-total density phase diagram.}
\end{figure}

For $\delta=0$ this gives an imaginary pole with $\alpha_0^+\geq\lambda$,
pertaining to the positive sign, and a `false' solution obtained with the
negative sign giving $\alpha_0=\lambda$ for all state points. The imaginary pole
is analogous to that found in the RPA solution for the OCY.
Fig.~\ref{fig:poles1} shows how this pole ascends from $\alpha_0=\lambda$ at
$\rho=0$ as the density $\rho$ is increased, along a line of constant
concentration, $x_2=0.5$, at fixed temperature $T^*=1$.

By calculating the residues one can show (see Eq.~(\ref{eq:rpaamp2}) below)
that the `false' pole $\alpha_0=\lambda$ makes no contribution to $h_{ij}(r)$,
i.e.~for $\delta=0$ the corresponding amplitude $A_{ij}^-=0$ and
\begin{equation}
rh_{ij}^{RPA}(r)=A_{ij}^+\exp(-\alpha_0^+r)
\label{eq:rparhrd0}
\end{equation}
where $\alpha_0^+$ is given by Eq.~(\ref{eq:rpasol}), with the positive sign, and
the amplitudes are independent of the density\cite{Note1}
\begin{equation}
A_{ij}^+=-\frac{M_{ij}}{\lambda T^*}.
\label{eq:rpaamp1}
\end{equation}

The situation is quite different for $\delta>0$. We find that the pole with
$\alpha_0^+>\lambda$ is modified slightly by the addition of a (relatively)
small positive term. More significantly we find that the solution with the
negative sign now corresponds to a second pole which descends from
$\alpha_0^-=\lambda$ eventually reaching zero as the density is increased -- see
Fig.~\ref{fig:poles2}. The locus of points in the phase diagram for which
$\alpha_0^-=0$ is the spinodal, where the $\hat{h}_{ij}(q)$ (or the partial
structure factors) diverge in the limit $q\rightarrow0$, indicating that the
system is undergoing fluid-fluid demixing. Note that this second `spinodal' pole
exists only for the mixture states, i.e.~for
$x_1,x_2\neq0$; in the pure states it
reverts to the `false' solution $\alpha_0=\lambda$. The behaviour of this pole
is shown for fixed $x_2=0.5$, and increasing density in Fig.~\ref{fig:poles2}. 

The `spinodal' pole is present for arbitrarily small positive $\delta$.
Therefore the RPA predicts that the system will undergo phase separation
provided there is any degree of positive
non-ideality in the pair potentials. This behaviour is reflected
in the asymptotic decay of the total correlation functions. For $\delta>0$,
\begin{equation}
rh_{ij}^{RPA}(r)=A_{ij}^+\exp(-\alpha_0^+r)+A_{ij}^-\exp(-\alpha_0^-r)
\label{eq:rparhrd1}
\end{equation}
with $\alpha_0^-<\lambda<\alpha_0^+$, which follows from Eq.~(\ref{eq:rpasol}).
The particularly simple forms of Eqs.~(\ref{eq:rparhrd0}) and
(\ref{eq:rparhrd1}) are a consequence of the RPA.\cite{Note1} The amplitudes are
\begin{equation}
A_{ij}^{\pm}=-\frac{1}{4\lambda T^*}
\frac{2M_{ij}\left(M_0\pm\sqrt{M_0^2+M_{\delta}}\right)
+\delta_{ij}M_{\delta}/x_i}{M_0+M_{\delta}/
\left(M_0\pm\sqrt{M_0^2+M_{\delta}}\right)},
\label{eq:rpaamp2}
\end{equation}
where superscript $\pm$ refers to the sign in front of the square root and we
use the
compound parameters $M_0$ and $M_{\delta}$ defined previously. $\delta_{ij}$ is
the Kronecker delta. As the ultimate decay is determined by the pole with the
smallest imaginary part, the `spinodal' pole $\alpha_0^-$ must determine the
ultimate decay of $h_{ij}(r)$ for any $\delta>0$. For very small $\delta>0$,
$\alpha_0^-$ lies very close to $\lambda$ until the reduced density
$\rho\lambda^{-3}$ reaches very large values, i.e.~the spinodal is shifted to
very large densities as $\delta \rightarrow 0$. In these circumstances the
relative magnitude of the amplitudes $A_{ij}^+$ and $A_{ij}^-$ becomes
important; note that $\alpha_0^+$ depends only
weakly on $\delta$. For a given state
point the amplitude of the `spinodal' pole $A_{ij}^-$ decreases relative to
$A_{ij}^+$ as $\delta\rightarrow0$ and one must go to increasing values of $r$
before the `spinodal' pole provides the dominant contribution to $h_{ij}(r)$. At
intermediate $r$ the contribution from $\alpha_0^+$ determines the decay
behaviour.

\subsection{Poles in the HNC}
\label{subsec:HNC_poles}

For the OCY the HNC predicts a crossover line in the $(\rho, T)$ plane
separating a region where the asymptotic decay of $h(r)$ is damped oscillatory
from that where it is monotonic. The crossover takes place via the coalescence
of two imaginary poles to form a complex pole as the density is increased at
fixed temperature $T$.\cite{Hopkins} The mechanism is equivalent to that
discussed first by Kirkwood in pioneering studies of strong
electrolytes.\cite{Kirkwood} Thus for the pure species we take the results of
Ref.~\onlinecite{Hopkins} and read off the crossover values of
$\rho\lambda^{-3}$. For $T^*=1$ and $x_2=0$ (pure species 1) crossover occurs at
$\rho\lambda^{-3}\approx0.47$ whereas for $x_2=1$ (pure species 2) this occurs
at $\rho\lambda^{-3}\approx0.025$. If we vary the concentration between these
limiting values (at fixed $T^*=1$) we expect to find a crossover line in the
$(x_2, \rho\lambda^{-3})$ plane. We determined this line by calculating the
poles of $\hat{h}_{ij}(q)$, i.e.~the zeros of $D(q)$, using the same numerical
procedure as in Ref.~\onlinecite{Hopkins} for the OCY.

The pair direct correlation functions $c_{ij}(r)$ are obtained from the full
solutions of the HNC integral equations. In order to ensure convergence of the
integrals which determine the poles we follow the procedure given in
Refs.~\onlinecite{Hopkins} and \onlinecite{LeoDC99}. For
$r\rightarrow\infty$, $c_{ij}(r)\rightarrow-M_{ij}\exp(-\lambda r)/(\lambda T^*
r)$, so we define a short range direct correlation function $c_{ij}^{sr}(r)$:
\begin{equation}
c_{ij}^{sr}(r)\equiv c_{ij}(r)+\frac{M_{ij}\exp(-\lambda r)}{\lambda T^*r}.
\label{eq:dcfsep1}
\end{equation}
Fourier transforming we find
\begin{equation}
\hat{c}_{ij}(q)=\hat{c}_{ij}^{sr}(q)
-\frac{4\pi M_{ij}}{\lambda T^{*}}\frac{1}{(q^{2}+\lambda^{2})},
\label{eq:dcfsep2}
\end{equation}
which can be substituted into Eq.~(\ref{eq:den}). $D(q)$ is separated into its
real and imaginary components, and the equation $D(q)=0$ is solved numerically
using a Newton-Raphson procedure. In general, the relevant integrals converge
only for complex $q$ such that $\Im[q]\lesssim2\alpha_0$, where $\alpha_0$ is
the imaginary part of the leading order pole.\cite{LeoDC99} It follows that
only a few poles can be calculated; the remaining poles are situated outside
this region of convergence. Fortunately the poles relevant for determining the
basic features of the decay of correlations can be obtained.

\begin{figure}
\includegraphics[width=8cm]{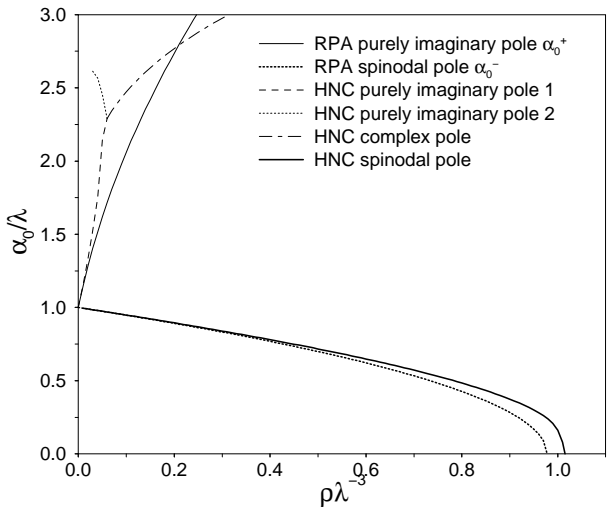}
\caption{\label{fig:poles2}
As in Fig.~\ref{fig:poles1} but now for $\delta=0.1$. Within the RPA the
imaginary pole $\alpha_0^+$, is shifted by only a small amount from the case
$\delta=0$. A second purely imaginary pole $\alpha_0^-$ is introduced which
decreases from $\lambda$ to zero as $\rho$ is increased;
$\alpha_0^-=0$ corresponds to the spinodal.
Since $\alpha_0^-<\alpha_0^+$ this `spinodal' pole determines the ultimate decay
of the total correlation functions. Within the HNC closure the Kirkwood
mechanism is still present and the crossover point is largely unchanged from the
case $\delta=0$. However, a `spinodal' pole is introduced which follows closely
the corresponding RPA pole $\alpha_0^-$. It follows that for {\it all}
densities the ultimate decay of the total correlation functions is monotonic.
Although the Kirkwood crossover is present it does not involve the leading
order `spinodal' pole and therefore does not influence the ultimate decay.}
\end{figure}

For the case $\delta=0$ we find that for all $x_2$ the leading
poles exhibit Kirkwood
cross-over mimicking that in the OCY.\cite{Hopkins} At very low densities we
find an imaginary pole (dashed line in Fig.~\ref{fig:poles1}) that ascends
from $\alpha_0=\lambda$ as we increase the density $\rho$ at constant
concentration, and is initially similar to the RPA. At slightly higher densities
a second imaginary pole (dotted line) moves into the region of convergence and
descends as $\rho$ increases. These two poles move towards each other and then,
at the Kirkwood point, coalesce. On increasing the density further, one finds a
conjugate complex pair whose real and imaginary parts both increase with $\rho$
(dash-dotted line). Fig.~\ref{fig:poles1} shows the imaginary parts of the
relevant poles as they undergo the Kirkwood cross-over. The locus of points for
which the poles coalesce, the Kirkwood line, is shown in the inset to
Fig.~\ref{fig:poles1}. Below this line the asymptotic decay of $rh_{ij}(r)$ is
pure exponential and above the line it is exponentially damped oscillatory. One
sees that the state point $x_2=0.5$ and $\rho\lambda^{-3}=0.5$, which
corresponds to the results for $h_{ij}(r)$ in Fig.~\ref{fig:rdfs}(a), lies deep
within the oscillatory region so one expects to find oscillations extending to
$r\rightarrow\infty$. 

Fig.~\ref{fig:poles2} displays the corresponding plots of $\alpha_0$ for
$\delta=0.1$. Now there are {\it three} purely imaginary poles at low densities.
Two of these mimic closely what is found for $\delta=0$, i.e, they coalesce and
form a complex pole at a Kirkwood point that is not far removed from the
corresponding point for $\delta=0$. Once again one of these poles (dashed line)
lies close to the RPA solution $\alpha_0^+$ for very small values of
$\rho\lambda^{-3}$. The third pole follows closely the RPA `spinodal' pole
$\alpha_0^-$; it decreases slowly with increasing $\rho\lambda^{-3}$ for small
densities before decreasing rapidly to zero at the spinodal point which is at
$\rho\lambda^{-3}\approx1.01$, slightly higher than the value
$\rho\lambda^{-3}=0.98$ found from the RPA. We refer to the third pole as the
HNC `spinodal' pole.

\begin{figure}
\includegraphics[width=8cm]{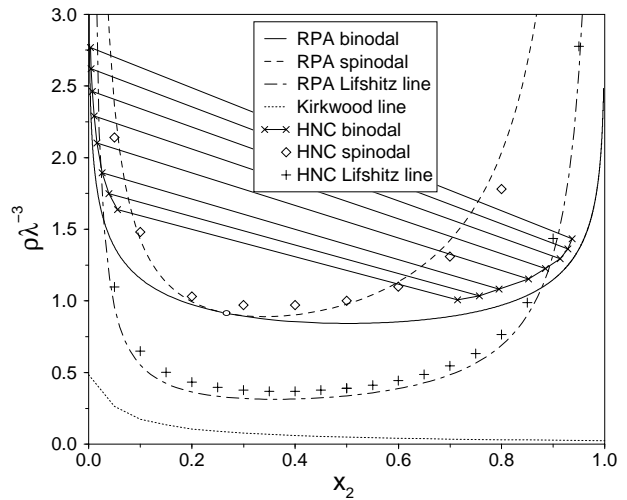}
\caption{\label{fig:phased}
Phase diagram for the binary Yukawa system with $M_{22}/M_{11}=4$, $T^*=1$ and
$\delta=0.1$. The system exhibits phase separation within both the RPA and HNC
closures. The RPA spinodal (dashed curve) meets the RPA binodal (solid curve) at
the critical point $\circ$. The HNC spinodal points (diamonds) were calculated
by extrapolating the `spinodal' pole to zero along lines of constant
concentration and increasing density. The HNC binodal was calculated using the
method outlined in the text; the straight segments are tie-lines connecting
coexisting fluid phases denoted by crosses and correspond to the following
reduced pressures $\beta P \lambda^{-3}=$ 20, 22, 25, 30, 35, 40, 45 and 50.
Also shown is the Lifshitz line for $S_{NN}(q)$ calculated in both the RPA
(dash-dotted) and HNC (crosses). The dotted line at low density denotes Kirkwood
crossover from monotonic to oscillatory decay of correlation functions. However
for $\delta>0$ and $0<x_2<1$ the true asymptotic decay remains monotonic even
for states above the line -- see text.}
\end{figure}

It is evident from Fig.~\ref{fig:poles2} that within the HNC, as in the RPA, the
purely imaginary `spinodal' pole with $\alpha_0<\lambda$ will determine the
ultimate decay of $h_{ij}(r)$ which will be monotonic for all densities provided
$0<x_2<1$. The ultimate decay cannot be oscillatory. Although there is still a
Kirkwood-like crossover, this now involves higher order poles rather than the
leading order poles, which
was the case for $\delta=0$. The corresponding crossover
line is shown as the dotted line in the $(x_2, \rho\lambda^{-3})$ phase diagram
for $\delta=0.1$ displayed in Fig.~\ref{fig:phased}. Note that the line lies far
below the HNC spinodal (diamonds). Crossover does not influence the ultimate
decay of $h_{ij}(r)$ which is stubbornly monotonic.

Of course, this does not mean that we do not find oscillations at intermediate
$r$. For densities larger than the crossover value the HNC pole analysis
predicts, for $r\rightarrow\infty$, 
\begin{equation}
rh_{ij}(r)\sim A_{ij}\exp(-\alpha_0r)
+2\tilde{A}_{ij}\exp(-\tilde{\alpha}_0r)\cos(\alpha_1r-\tilde{\theta}_{ij})
\label{eq:rhrhnc}
\end{equation}
with $\alpha_0<\tilde{\alpha}_0$. Provided $\tilde{A}_{ij}\gg A_{ij}$ and we are
not close to the spinodal ($\alpha_0=0$) the oscillatory contribution will be
significant in the intermediate regime.

\begin{figure}
\includegraphics[width=8cm]{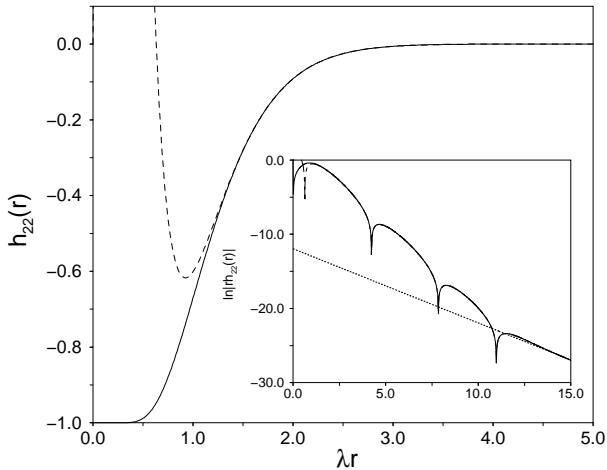}
\caption{\label{fig:hrcomp}
Comparison between results of the full HNC solution for $h_{22}(r)$ (solid line)
and the contributions of the lowest order poles for state point
$\rho\lambda^{-3}=0.05$, $x_2=0.99$ and parameters $M_{22}/M_{11}=4$, $T^*=1$
and $\delta=0.1$. Note this point lies above the Kirkwood crossover line in
Fig.~\ref{fig:phased}. The main figure shows that the decay of $h_{22}(r)$
is well described by the contribution from a single pair of
conjugate complex poles (dashed line) ($\tilde{\alpha}_0/\lambda_0=2.269,$
$\alpha_1/\lambda=0.8723$). On this scale the contribution from the purely
imaginary pole is vanishingly small. In the inset the plot of $\ln|rh_{22}(r)|$
versus $\lambda r$ (solid line) shows that although the oscillatory
contribution (dashed line -- visible only in top left hand corner) describes
$h_{22}(r)$ accurately for intermediate range, the contribution from the
`spinodal' pole (dotted line) ($\alpha_0/\lambda=0.9993$)
determines the decay for
$\lambda r\gtrsim14$. Thus the sum of contributions from the oscillatory and
`spinodal' poles (dot-dashed) given by Eq.~(\ref{eq:rhrhnc}) is very accurate in
the range $1.5<\lambda r<\infty$.}
\end{figure}

An example is given in Fig.~\ref{fig:hrcomp} which compares the full numerical
result for $h_{22}(r)$ with the contribution given by Eq.~(\ref{eq:rhrhnc}) for
a statepoint $(x_2=0.99, \rho\lambda^{-3}=0.05)$ where the amplitude of the
`spinodal' pole is very small. On a normal scale plot $h_{22}(r)$ is
well-described by the
contribution from a single conjugate complex pair of poles, i.e.~by the second
term in Eq.~(\ref{eq:rhrhnc}). This contribution is oscillatory, albeit with a
small amplitude. The inset to Fig.~\ref{fig:hrcomp} plots
$\ln|rh_{22}(r)|$ for a larger range of $\lambda r$. Now the oscillations are
manifest but we observe monotonic (exponential) decay, described accurately
by the first term in Eq.~(\ref{eq:rhrhnc}), for $\lambda r\gtrsim14$ (dotted
line). As the amplitude $\lambda A_{22}$ of the `spinodal' pole contribution is
$\sim10^{-6}$ this contribution does not become significant until large
separations $r$. Of course such a contribution might be impossible to detect
experimentally or in simulations.
We conclude that although a pair correlation function might
exhibit many oscillations, the ultimate decay can still be monotonic. Note that
for a state point closer to the spinodal, such as in Fig.~\ref{fig:rdfs}(c)
where $x_2=0.5$, $\rho\lambda^{-3}=0.5$ and $\alpha_0/\lambda\approx0.72$, the
purely exponential contribution in Eq.~(\ref{eq:rhrhnc}) dominates for $\lambda
r \gtrsim 3$.

\begin{figure}
\includegraphics[width=8cm]{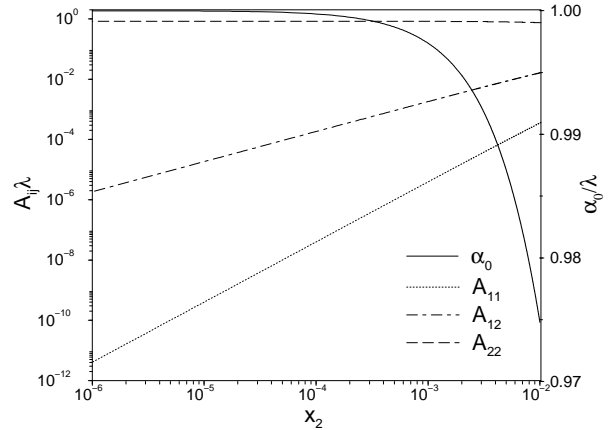}
\caption{\label{fig:polenx0}
The HNC `spinodal' pole $\alpha_0$ and corresponding amplitudes $A_{ij}$ as a
function of concentration $x_2$ along a path of constant density
$\rho\lambda^{-3}=0.5$, for $T^*=1$, and $\delta=0.1$.  As $x_2\rightarrow 0$,
$\alpha_0/\lambda\rightarrow1^-$, $A_{11}, A_{12}\rightarrow0$ but
$A_{22}\lambda$ tends to 0.84. Note that $A_{12}$ is negative and its magnitude
is plotted here. The amplitudes obey the rule $A_{12}^2=A_{11}A_{22}$ -- see
text.}
\end{figure}

For a pure (one component) fluid there is no `spinodal' pole; within the HNC the
Kirkwood crossover occurs between leading-order poles. Therefore, on the axes
$x_2=0$ and $x_1=0$ in Fig.~\ref{fig:phased} $h_{ii}(r)$ must decay in an
oscillatory fashion as $r\rightarrow\infty$, provided we consider states above
the Kirkwood crossover densities. However, as we increase $x_2$ (or $x_1$)
infinitesimally the `spinodal' pole makes a non--vanishing contribution to the
correlation functions and forces the ultimate decay of $h_{ij}(r)$ to be
monotonic. In order to understand the evolution of this behaviour we must
consider the amplitudes, $A_{ij}$, of the `spinodal' pole as the concentration
$x_2$ vanishes along a line of constant total density. Fig.~\ref{fig:polenx0}
shows results for $\rho\lambda^{-3}=0.5$, a density that lies just above the
Kirkwood crossover value, $\rho\lambda^{-3}\sim0.47$, for the pure fluid. The
`spinodal' pole $\alpha_0/\lambda\rightarrow1^-$ as $x_2\rightarrow0$, and the
accompanying amplitudes $A_{11}$ (positive) and $A_{12}$ (negative) both vanish,
as power laws,
in this limit. It follows that the `spinodal' pole makes contributions to
$h_{11}(r)$ and $h_{12}(r)$ that become vanishingly small in a continuous
fashion, as $x_2\rightarrow0$. Thus for very small concentrations $h_{11}(r)$
and $h_{12}(r)$ appear oscillatory until extremely large values of $\lambda r$
where the `spinodal' pole will dictate the final (monotonic) decay of these
functions.

The amplitude $A_{22}$ of the minority component, species 2, has a different
variation. As shown in Fig.~\ref{fig:polenx0}, $\lambda A_{22}$ remains almost
constant as the concentration is reduced and tends to a non-zero value (0.84) at
$x_{2}=0$. At first sight this result is a little surprising. However we should
recall that physical observables such as probability distributions or the liquid
structure factors involve the product $\rho_2h_{22}(r)$ which vanishes as
$x_2\rightarrow0$. We confirmed that our numerical results for
the amplitudes satisfy the rule\cite{Evans} $A_{12}^2=A_{11}A_{22}$ mentioned
in Sec.~\ref{sec:asymp_decay}.

At this point it is instructive to return to the RPA `spinodal' pole. Recall
that $\alpha_0^-$ is given by Eq.~(\ref{eq:rpasol}) and the corresponding
amplitudes by Eq.~(\ref{eq:rpaamp2}). If we Taylor expand around $x_2=0$ we
find that
\begin{equation}
\alpha_0^-=\lambda \left(1-\frac{2 \pi \rho M_{22} \delta (2+\delta)}
{\lambda^3 T^*} x_2\right)+O(x_2^2),
\label{eq:taylor_alpha}
\end{equation}
$A_{11}^-$ and $A_{12}^-$ decay to zero as power laws in $x_2$, and
that $A_{22}^-$ tends to a constant, i.e.
\begin{eqnarray}
A_{11}^-&=&\frac{M_{22}^2(1+\delta)^2(2+\delta)\delta}
{\lambda T^* M_{11}} x_2^2 + O(x_2^3)\nonumber,\\
A_{12}^-&=&-\frac{M_{22}^{3/2}(1+\delta)(2+\delta)\delta}
{\lambda T^* M_{11}^{1/2}}x_2 +O(x_2^2),\label{eq:rpa_ampnx0}\\
A_{22}^-&=&\frac{M_{22}(2+\delta)\delta}{\lambda T^*} +O(x_2)\nonumber.
\end{eqnarray}
In the limit $x_2\rightarrow0$, $\lambda
A_{22}^-=M_{22}(2+\delta)\delta/T^*$ which takes the value $0.84$ for the
present choice of parameters: $M_{22}=4$, $\delta=0.1$ and $T^*=1$. This is the
same limiting value of the amplitudes as obtained in the HNC. Indeed we find
that the RPA results for $\alpha_0^-$ and for the amplitude $A_{ij}^-$
are almost identical to those from the numerical HNC results for concentrations
$x_2\lesssim10^{-4}$. Thus, numerically, the limiting behaviour predicted by
Eqs.~(\ref{eq:taylor_alpha}) and (\ref{eq:rpa_ampnx0})
pertains to the HNC as well as to the RPA. Note that the
coefficients of the leading order terms in Eq.~(\ref{eq:rpa_ampnx0}) are
consistent with the requirement $(A_{12}^-)^2=A_{11}^-A_{22}^-$.

It is not immediately obvious why the two (very) different closure approximations
should yield the same limiting behaviour for the pair correlation functions as
$x_2 \rightarrow 0$. That they do suggests that the limiting behaviour described
by Eqs.~(\ref{eq:taylor_alpha}) and (\ref{eq:rpa_ampnx0}) should be valid more
generally. In the Appendix we show that these results follow from the natural
division, Eq.~(\ref{eq:dcfsep1}), of the pair direct correlation functions into
a Yukawa tail plus a short ranged contribution $c_{ij}^{sr}(r)$. Provided the
Fourier transforms $\hat{c}_{ij}^{sr}(q)$ are sufficiently well--behaved in the
neighbourhood of $q=i\lambda$, Eqs.~(\ref{eq:taylor_alpha}) and
(\ref{eq:rpa_ampnx0}) should be valid for {\em any} closure.

\section{Phase Diagram and Lifshitz line for $\delta=0.1$}
\label{sec:phase_diag}

The results presented in the last section imply that within the RPA and HNC
introducing a small degree of non-ideality into the pair potentials defining the
mixture, i.e.~imposing $\delta>0$, leads to a `spinodal' pole which dictates
that pair correlation functions should decay monotonically as
$r\rightarrow\infty$ for {\em all} state points, including those far removed
from the spinodal, where one might have
expected oscillatory decay (within the HNC). In this section we focus on the
spinodal itself and the associated fluid-fluid phase separation.

The spinodal is the locus of points in the phase diagram at which the `spinodal'
pole reaches zero. Within the RPA we can determine the spinodal by finding
solutions of $\alpha_0^-=0$, where $\alpha_0^-$ is given by
Eq.~(\ref{eq:rpasol}) with the negative sign. This result is plotted in
Fig.~\ref{fig:phased} for $\delta=0.1$ and $T^*=1$ (dashed curve). Within the
HNC it is not possible to determine the spinodal precisely. For certain
densities and concentrations the HNC approximation does not have solutions.
However it is possible to calculate the position of the HNC `spinodal' pole as a
function of $\rho\lambda^{-3}$ at fixed concentration and extrapolate to zero as
shown in Fig.~\ref{fig:poles2}. These spinodal points, obtained for a number of
concentrations, are shown in Fig.~\ref{fig:phased} (diamonds). The HNC spinodal
lies close to that obtained in the RPA. 

Using the RPA we are able to write the reduced bulk Helmholtz free energy per
particle, $\tilde{f}$, as a sum of ideal and excess parts\cite{Louis, Archer02,
Archer01, Finken}
\begin{equation}
\tilde{f}(\rho,x_2)=\tilde{f}_{id}
+\frac{1}{2}\rho[x_1^2\hat{\phi}_{11}(0)+2x_1x_2\hat{\phi}_{12}(0)
+x_2^2\hat{\phi}_{22}(0)]
\label{eq:helmfree}
\end{equation}
where $\hat{\phi}_{ij}(0)=4\pi\epsilon M_{ij}\lambda^{-3}$ is the $q=0$ limit of
the Fourier transformed pair potential. The ideal part,
$\tilde{f}_{id}(\rho,x_2)$, contains the ideal free energy of mixing,
$\beta^{-1}\{x_1\ln(x_1)+x_2\ln(x_2)\}$, as well as a term in $\rho$ that is
irrelevant for phase behaviour. Eq.~(\ref{eq:helmfree}) corresponds to
calculating $\tilde{f}$ from the compressibility route, and the spinodal
obtained from this equation is identical to that obtained from the zeros of
$\alpha_0^-$. We now Legendre transform to the Gibbs free energy
$g=\tilde{f}+Pv$, where $v=1/\rho$ is the volume per particle and the pressure
is given by $P=-(\partial\tilde{f}/\partial v)_{x_2}$. Using the common tangent
construction on $g$ we obtain the binodal which is also plotted in
Fig.~\ref{fig:phased} (solid curve). The binodal and spinodal meet at the
critical point near $x_2=0.27$, $\rho\lambda^{-3}=0.88$.

In implementing the HNC we choose to calculate the thermodynamic functions
locally, thereby avoiding thermodynamic integration. We begin by tracing out
isobars across the phase diagram with the pressure given by the virial route:
\begin{equation}
\frac{\beta P}{\rho}
=1-\frac{2\pi\rho}{3}\sum_{i}\sum_{j}x_ix_j\int_{0}^{\infty}
\dd r r^3 \frac{\dd \beta \phi_{ij}(r)}{\dd r} g_{ij}(r).
\label{eq:presvir}
\end{equation}
Along the isobars we calculate the HNC chemical potentials
\begin{eqnarray}
\beta\mu_i&=&\ln(\rho_i\Lambda_i^3)\\
&&+\sum_j{\rho_j}\int \dr \left\{\frac{h_{ij}(r)}{2}[h_{ij}(r)-c_{ij}(r)]
-{c}_{ij}(r)\right\}\nonumber
\label{eq:chempot}
\end{eqnarray}
where $\Lambda_i$ is the (irrelevant) thermal de Broglie wavelength of species
$i$, with $i=1,2$. The Gibbs free energy per particle is then given by
$g=x_1\mu_1+x_2\mu_2$. By fitting a polynomial to these results we are then able
to determine the binodal using the common tangent construction. For the isobars
that intersect the region for which the HNC does not provide a solution we
calculate the two `branches' of the free energy and construct the common tangent
on these; the procedure is equivalent to that used in
Ref.~\onlinecite{Archer02}.
Nevertheless, for state points close to the critical point it is not possible to
determine the binodal. The HNC binodal is shown in Fig.~\ref{fig:phased}
(crosses connected by tie lines). This lies outside our estimated HNC spinodal
and fairly close to the RPA binodal. A slightly better level of agreement
between the HNC and RPA results for the binodal was found in
Ref.~\onlinecite{Archer02} for a softcore model of binary star-polymers.

Since the `spinodal' pole dictates the ultimate decay of the pair correlations
even for states that are far removed from the spinodal it is instructive to seek
some criterion which indicates when contributions from the other pole(s) become
important in determining the structure of the fluid. One valuable indicator of
the change in the latter is the Lifshitz line which focuses on the behaviour of
the fluid structure at small wave numbers $q$. \cite{Archer02, Gompper} We
define the partial structure factors as\cite{Caccamo}
\begin{equation}
S_{ij}(q)=\delta_{ij}+(x_ix_j)^{1/2}\rho\hat{h}_{ij}(q),
\label{eq:sijq}
\end{equation}
and concentrate on the number-number structure factor 
\begin{equation}
S_{NN}(q)=x_1S_{11}(q)+x_2S_{22}(q)+2(x_1x_2)^{1/2}S_{12}(q).
\label{eq:snnq}
\end{equation}
As we approach the critical point or, indeed, the spinodal, the partial
structure factors diverge at small $q$: $S_{11}(q=0)$ and $S_{22}(q=0)
\rightarrow +\infty$ and $S_{12}(q=0)\rightarrow -\infty$. These divergences are
reflected in the linear combination, i.e.~$S_{NN}(q=0) \rightarrow +\infty$.

We define the Lifshitz line as separating regions of the phase diagram for which
$S_{NN}(q)$ has a local maximum or a local minimum at $q=0$. We choose
$S_{NN}(q)$ because i) it is the most symmetrical combination of the partial
structure factors, and ii) it diverges in a similar manner on both sides of the
phase diagram. Formally we make a small $q$ expansion of $S_{NN}(q)$:
\begin{equation}
S_{NN}(q)=a(\rho,x_2)+b(\rho,x_2)q^{2}+O(q^{4}),
\label{eq:lscr}
\end{equation}
so that the Lifshitz line is the locus of points for which $b(\rho,x)=0$.
\cite{Archer02, Gompper} This line is calculated in both the RPA and HNC and
the results shown as the dash-dotted curve and crosses, respectively, in
Fig.~\ref{fig:phased}. We find that there is good agreement between the two
closures, suggesting that in this region of the phase diagram the RPA should
become a fairly reliable approximation for describing $h_{ij}(r)$ at both
intermediate and large values of $r$.

\section{Discussion and Conclusions}
\label{sec:discussion}

In this paper we have investigated the structural and thermodynamic properties
of binary mixtures of particles interacting via purely repulsive (point)
Yukawa pair potentials.
We have used two approximate closures to the OZ equations in order to
calculate the fluid properties: the simple RPA allows us to elucidate
analytically some important properties, and the HNC approximation is
expected to be very accurate for the Yukawa fluid at the
intermediate fluid densities relevant to the present work. This expectation
stems from the fact that the HNC closure correlation functions are
rather accurate for the OCY.\cite{Hopkins,Daughton}

For the `ideal' mixture, with $\delta=0$, the RPA approximation gives a poor
description of the fluid structure, predicting
monotonic (exponential) decay of $rh_{ij}(r)$ for all state points in the phase
diagram rather than the Kirkwood crossover line that is manifest in the HNC --
see Fig.~\ref{fig:poles1}. No fluid--fluid phase separation occurs for
$\delta=0$. By contrast, when $\delta>0$ an additional purely imaginary
pole arises within the HNC approximation for $\hat{h}_{ij}(q)$. This
`spinodal' pole governs the phase separation in the mixture for the non--ideal
case. Since the RPA provides a good account of this pole -- see
Fig.~\ref{fig:poles2} -- it follows that the RPA also yields fluid-fluid
transition boundaries that are quite close to those from the more accurate HNC
-- see Fig.~\ref{fig:phased}. The amplitude $A_{ij}^-$ of the spinodal
pole contribution to $h_{ij}(r)$ becomes increasingly large with increasing
proximity to the spinodal, and this pole provides the dominant contribution in
the region of the phase diagram near the
spinodal (roughly the region enclosed by the Lifshitz line -- see
Fig.~\ref{fig:phased}).

For states away from the spinodal/binodal, particularly in the limits
$x_1 \rightarrow 0$
or $x_2 \rightarrow 0$, the amplitudes of the contribution from the
spinodal pole to the majority species correlation function $h_{ii}(r)$ (where
$i$ is the majority species) and to $h_{12}(r)$, become very small -- see
Figs.~\ref{fig:hrcomp} and \ref{fig:polenx0}. One must magnify the large $r$
tail of these correlation functions in order
to see the contribution from the spinodal
pole. Above the Kirkwood line, but well away from the spinodal, the
intermediate $r$ decay of $h_{ij}(r)$ is damped oscillatory; the oscillations
can persist to large $r$, before the monotonic decay from the spinodal
pole finally dominates at very large values of $r$. This result is quite
unusual, since one generally expects the leading order pole to
dominate the decay of $h_{ij}(r)$ at
both large {\em and} intermediate values of $r$. \cite{Hopkins,Evans,LeoDC99}
In the limit $x_2 \rightarrow 0$, the spinodal pole $\rightarrow i \lambda$
and the amplitudes, $A_{11}^-$ and $A_{12}^-$, of the contribution of to
$h_{11}(r)$ and to $h_{12}(r)$ have a power law dependence on
$x_2$ (see Eq.~(\ref{eq:rpa_ampnx0})), vanishing in
the pure fluid of species 1. However, the amplitude, $A_{22}^-$,
of the contribution from the spinodal pole to $h_{22}(r)$ tends to a
non--vanishing value in the limit $x_2
\rightarrow 0$. This means that for small values of $x_2$ the nature of the
decay of the correlation function $h_{22}(r)$ can be very different from that of
$h_{11}(r)$ and $h_{12}(r)$, a situation which has not been encountered in other
fluid mixtures. In experimental or simulation results it would be almost
impossible to detect the contribution in $h_{11}(r)$ and $h_{12}(r)$ from the
`spinodal' pole for small $x_2$.
However, its existence could be inferred from the apparently different decay of
$h_{22}(r)$ versus that of $h_{11}(r)$ and $h_{12}(r)$.

It is important to emphasise that
the amplitude $A_{22}^-$ tending to a non-zero value in the limit
$x_2 \rightarrow 0$ does not imply any pathological consequences for
thermodynamic or measurable structural quantities.
As pointed out earlier, physical observables such as the liquid
structure factors involve the product $\rho_2 h_{22}(r)$ which vanishes as
$x_2\rightarrow0$. One exception to this rule is the effective
potential between two particles immersed in a solvent of the other species.
The effective potential $\phi_{22}^{eff}(r)$ between two particles of species 2
at infinite dilution is related to
the radial distribution function, in the limit $x_{2}\rightarrow0$, by
\begin{equation}
g_{22}(r)\equiv \exp(-\beta\phi_{22}^{eff}(r)).
\label{eq:smp1}
\end{equation}
Clearly $\phi_{22}^{eff}(r)$ depends on the density and temperature of the
solvent (species 1). We define the solvent-mediated potential $W_{22}(r)$ via
$\phi_{22}^{eff}(r)\equiv\phi_{22}(r)+W_{22}(r)$, where $\phi_{22}(r)$ is the
bare (direct) potential between the two particles. $W_{22}(r)$ depends on the
nature of the solvent and on the solvent-particle interaction. From
Eq.~(\ref{eq:smp1}) it follows that $W_{22}(r)$ depends on $h_{22}(r)$ rather
than $\rho_{2}h_{22}(r)$. There are clear implications for the form of
$\phi_{22}^{eff}(r)$ given that the amplitude of the contribution to
$h_{22}(r)$ from the spinodal pole is non--vanishing in the limit
$x_2\rightarrow 0$. We leave a full discussion to another publication but point
out that the presence of the spinodal pole, which occurs for any $\delta>0$,
gives rise to an effective potential that is {\em attractive} at sufficiently
large distances and decays as $\exp(-\lambda r)/r$, i.e.~with the same length
scale as the bare potentials.

Given that the Yukawa potential arises in a great variety of physical
problems,\cite{rowlinson} typically where there are big charged particles
screened by a neutralising background medium, we believe the present work should
be relevant to systems ranging from charged colloidal fluids\cite{Hansen} to
dusty plasmas.\cite{piel:melzer2002}

We have chosen to focus on a simple example of a binary Yukawa mixture, namely
one in which the inverse screening length $\lambda$ is the same for all $ij$
pairs, as this is the situation which arises naturally for charged particles
immersed in a neutralising medium. One could, of course, consider more complex
examples with $\lambda$ dependent on the species indices $i$,$j$. Moreover, one
could consider mixtures that correspond to {\em negative} non--additivity,
i.e.~with $\delta<0$. In this case $M_{12}<\sqrt{M_{11}M_{22}}$ and the
interactions should favour fluid--fluid mixing.

\section*{Acknowledgements}
PH is grateful for the support of an EPSRC studentship and AJA 
acknowledges the support of EPSRC under grant number GR/S28631/01.

\section*{Appendix: Behaviour of correlation functions as $x_2 \rightarrow 0$}

Here we show that the results for the spinodal pole and the amplitudes
$A_{ij}^-$,
Eqs.~(\ref{eq:taylor_alpha}) and (\ref{eq:rpa_ampnx0}), in the limit
$x_2\rightarrow 0$ are not specific to the RPA but follow generally from the
form of the direct correlation functions for Yukawa mixtures. Making
the separation of the $c_{ij}(r)$ given by Eq.~(\ref{eq:dcfsep1}) and
substituting into Eq.~(\ref{eq:den}) we obtain
\begin{equation}
D(q)=a+\frac{b}{p}+\frac{c}{p^2},
\label{eq:D_app}
\end{equation}
where
\begin{eqnarray}
p&=&q^2+\lambda^2,
\nonumber \\
a&=&[1-\rho_1 c_{11}^{sr}(q)][1-\rho_2 c_{22}^{sr}(q)]
-\rho_1\rho_2[c_{12}^{sr}(q)]^2
\nonumber \\
b&=&[1-\rho_1 c_{11}^{sr}(q)]\rho_2 \alpha_{22}
+[1-\rho_2 c_{22}^{sr}(q)]\rho_1 \alpha_{11}\nonumber \\
&&+2\rho_1\rho_2c_{12}^{sr}(q) \alpha_{12},
\nonumber \\
c&=&\rho_1\rho_2(\alpha_{11}\alpha_{22}-\alpha_{12}^2)
\end{eqnarray}
and $\alpha_{ij}=4\pi M_{ij}/\lambda T^*$.
The poles are given by the solution to the equation $D(q)=0$. From equation
(\ref{eq:D_app}), we see that one set of solutions is given by
\begin{equation}
p_{\pm}=-(b \pm \sqrt{b^2-4ac})/2a.
\label{eq:p_appendix}
\end{equation}
$a$ and $b$ are functions of $q$. However, Eq.~(\ref{eq:p_appendix})
leads to purely imaginary poles at $q_{\pm}=i \alpha_0=i \sqrt{\lambda^2 -
p_{\pm}}$, provided we assume that the functions $c_{ij}^{sr}(q)$ are well
behaved (finite and differentiable) on the imaginary axis around $q_{\pm}$.
The leading order pole is that corresponding to $p_{-}$,
which in the limit of vanishing density $\rho_2=0$ (i.e.~$c=0$) yields a
pole at $q=i\lambda$, giving a decay $rh_{ij}(r) \sim A_{ij}^- \exp(-\lambda
r)$, where the amplitudes $A_{ij}^-$ are to be determined below. For small
concentrations of species 2 we can Taylor expand $p_-$ in powers of $c$, giving
\begin{equation}
p_-=-\frac{c}{\rho_1 \alpha_{11}}+\left( \frac{c (b-\rho_1\alpha_{11})}{\rho_1^2
\alpha_{11}^2}-\frac{ac^2}{\rho_1^3 \alpha_{11}^3} \right) +O(\rho_2^3).
\label{eq:p_expansion}
\end{equation}
We find that in the limit $\rho_2 \rightarrow 0$, $p_- \sim
-(\alpha_{22}-\alpha_{12}^2/\alpha_{11})\rho_2$. Thus the leading order pole
has
\begin{equation}
\alpha_0^-=\lambda \left(1-\frac{\alpha_{11}\alpha_{22}-\alpha_{12}^2}
{2\lambda^2 \alpha_{11}} \rho_2 \right)+O(\rho_2^2),
\label{eq:taylor_alpha_appendix}
\end{equation}
which is identical to the RPA result in Eq.~(\ref{eq:taylor_alpha}).

The amplitude, $A_{ij}^-$ is given by:
\begin{equation}
A_{ij}^-=\frac{q_- N_{ij}(q_-)}{2 \pi D'(q_-)},
\label{eq:amplitude}
\end{equation}
where the prime denotes the derivative with respect to $q$ and the
functions $N_{ij}(q)$ are given in Eq.~(\ref{eq:oznum}).
From Eq.~(\ref{eq:D_app}) we find that
\begin{equation}
D'(q)=a'+\frac{b'}{p}-\frac{2qb}{p^2}-\frac{4qc}{p^3}.
\label{eq:D_app_prime}
\end{equation}
In the limit $\rho_2 \rightarrow 0$, using (\ref{eq:p_expansion}), we find that
$D'(q_-)=2q_-b/p_-^2+$other terms less singular in $\rho_2$. Using this result,
we can expand in Eq.~(\ref{eq:amplitude}) to obtain the amplitudes $A_{ij}^-$ in
the limit $\rho_2 \rightarrow 0$. We find that the leading order terms are
precisely the RPA results given in Eq.~(\ref{eq:rpa_ampnx0}) -- i.e.~ in the
limit $\rho_2\rightarrow 0$ the amplitudes $A_{ij}^-$ are independent of the
functions $\hat{c}_{ij}^{sr}(q)$.

\end{document}